# Calculating the free energy of 2D materials on substrates

Yu-Peng Liu[1,2], Bo-Yuan Ning[3], Le-Cheng Gong[1,2], Tsu-Chien Weng[3], and Xi-Jing Ning[1,2,*]

[1]Institute of Modern Physics, Fudan University, Shanghai 200433, China
[2]Applied Ion Beam Physics Laboratory, Fudan University, Shanghai 200433, China
[3]Center for High Pressure Science& Technology Advanced Research, Shanghai, 202103, China

## Abstract

A method was developed to calculate the free energy of 2D materials on substrates and was demonstrated by the system of graphene and γ-graphyne on copper substrate. The method works at least 3 orders faster than state-of-the-art algorithms, and the accuracy was tested by molecular dynamics simulations, showing that the precision for calculations of the internal energy achieves up to 0.03% in a temperature range from 100K to 1300K. As expected, the calculated the free energy of a graphene sheet on Cu (111) or Ni (111) surface in a temperature range up to 3000K is always smaller than the one of a γ-graphyne sheet with the same number of C atoms, which is consistent with the fact that growth of graphene on the substrates is much easier than γ-graphyne.

*Correspondence should be addressed to xjning@fudan.edu.cn

# 1. Introduction

Since graphene was prepared successfully in 2004 [1], 2-dimensional (2D) materials have developed a wide interest all over the world. Due to the ultrathin thickness at the atomic level, the materials exhibit unique electrical, mechanical and thermal properties, which inspires researchers to explore and discover other more 2D materials. So far, dozens of 2D materials have been prepared experimentally, including graphene family (e.g., graphdiyne[2], silicene[3], germanene[4], BN[5], borophene[6], Phosphorene[7], bismuthene[8]), transition metal dichalcogenides [9], metal carbides[10], and the like, but it yet remains a problem to prepare high quality 2D material with larger sizes. Although a few kinds of 2D materials such as graphene and $MoS_2$[11] can be obtained by mechanical exfoliation of the corresponding bulk materials, vapor deposition (VD) such as chemical vapor deposition (CVD) should be much more flexible for preparing various 2D materials of large scale[12], which can be easily transferred to other substrates. However, this method requires a strict growth condition because the surface structure and the temperature of substrate both have a significantly effect on the growth of 2D materials, and lots of time and effort have to be payed to explore the growth conditions through continuous experiments. For example, deposit of silicon atoms on Ag(001)[13] and Ag(110)[14] surface can only produce Si superstructure or one-dimensional silicene nanoribbons, while a continuous film of honeycomb structure, silicene[3, 15], can be obtained once the substrate was replaced with Ag(111). Even in such a case, it yet remains uncertain what the most optimum conditions are (maybe another substrate heated at different temperatures) for silicene growth. The solution may resort to theoretical predications instead of endless experimental exploration.

In previous theoretical exploration, the interaction potential energy between 2D materials and substrates were calculated to see effects of the surface structures on formation of 2D materials. For example, for the growth of graphyne, Crljen et al. studied the adsorption energy of graphyne on Cu (111), Ni (111) and Co (0001) surfaces [16], Ding et al. studied the geometric structure of carbyne on various transition metal surfaces through the formation energy[17], and Yang et al. investigated the interfacial structural of graphdiyne on Cu (111) surface through bidding energy[18]. These work provides useful information on the effect of substrates, while it can't take into account the effects of temperature.

In principle, calculations of the free energy can predict which surface structure of a substrate at what temperature is more favorable to the growth of given 2D materials by VD. However, it has been an open problem since the born of statistical physics by the end of $19^{th}$ century to calculate the free energy of condensed matters. In the past 30 years, substantial progress has been made in calculations of partition function (PF) for condensed matters, from which the free energy as well as other thermodynamic state functions can be obtained. Among the advanced methods including parallel tempering[19], umbrella sampling[20], metadynamics[21], Wang-Landau sampling[22], and Nested sampling[23, 24], an algorithm of Nested sampling (ANS) developed by [24] may be state of the art and can produce the PF for condensed system composed of hundreds of atom if empirical pairwise potential is applied for the interaction of atoms. In the algorithm of ANS, all the atoms in the system should be moved artificially in real space so as to produce enough configurations (usually more than $10^4$) to produce reliable PF. When the algorithm is applied to calculate the PF of a given 2D material, such as a graphene sheet with perfect hexagons structure, artificially moving the C atoms in real space may produce structures approaching to the ones of graphyne or others, so the final obtained PF may not related uniquely with the graphene sheet. Clearly, artificial constrains must be applied to moving the atoms so as the produced molecular configurations are closely related to the structure of the given 2D materials, which will result in large uncertainty of the PF. In such case, the difference of the free energy derived from the PF between two different 2D structures will depend too much on the artificial constrains, leading to failure of the free energy criterion for predicting which 2D structure should be more favorable.

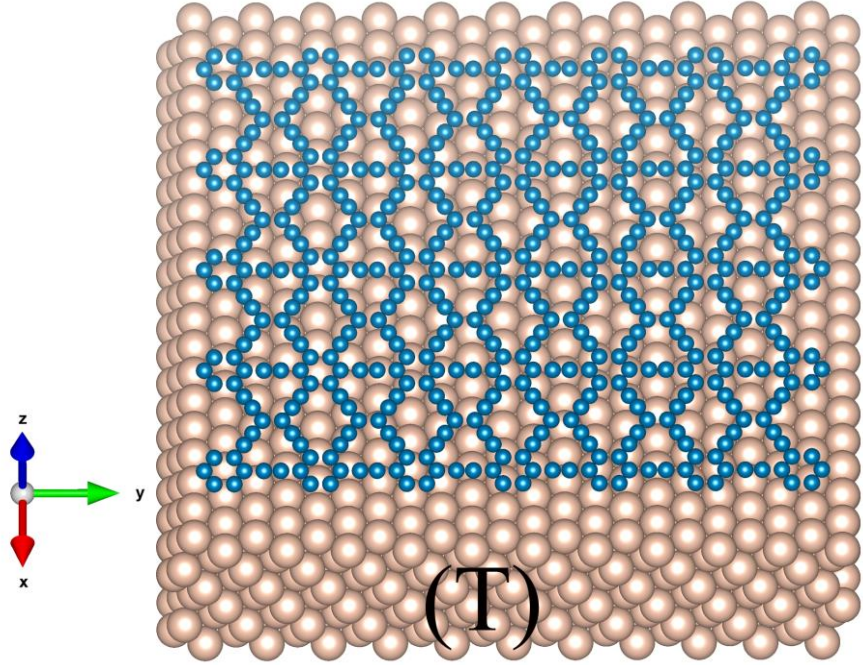

**Figure 1.** Schematic of a piece of 2D material of *N* atoms (red balls) lying on the surface of a substrate of *M* atoms (golden balls) at temperature *T*.

In the present work, a direct integral approaches[25] was developed to calculate the PF of 2D materials and was demonstrated by 2D carbon materials on Cu substrate. The high calculation efficiency enables us to obtain the PF of a graphene sheet composed of more than 500 atoms with Brenner Potential[26](a many-body interaction function) by using a personal computer in about an hour. In order to test the accuracy, we performed molecular dynamics (MD) simulations of a graphene (or γ-graphyne) sheet of 510 atoms on Cu (111) surface at temperatures up to 1300K to produce the internal energies to be compared with the ones derived from the PF using the same interaction potential, showing that the relative deviations is smaller than 0.03%. The free energy of graphene derived from the PF is always smaller than the one of γ-graphyne in the temperature range of 0-3000K, which is in agreement with the fact that growth of graphene under such conditions is much easier than γ-graphyne.

## 2. theoretical method

The model for a 2D materials on a substrate is shown in Fig.1, where the substrate of $M$ atoms is treated as a thermal bath at temperature $T$. For calculations of the PF for the 2D materials of $N$ atoms, the total potential is expressed as

$$U(x^{3N}, X^{3M}) = U_{2D}(x^{3N}) + V(x^{3N}, X^{3M}), \qquad (1)$$

where $U_{2D}$ is the potential energy of the 2D material with the coordinates of the atoms denoted by $x^{3N}(x_1, x_2 ... x_{3N})$, and $V$ is the interaction potential between the 2D material and the substrate with its atoms denoted by $X^{3M}(X_1, X_2 ... X_{3M})$.

The PF of 2D materials can be expressed as,

$$Z = \frac{1}{N!}\left(\frac{2\pi m}{\beta h^2}\right)^{\frac{3N}{2}} Q, \qquad (2)$$

where $\beta = 1/k_B T$ with $k_B$ the Boltzmann factor, and $Q$ is the configurational integral

$$Q = \int d x^{3N} \exp[-\beta U(x^{3N}, X^{3M})]. \qquad (3)$$

In order to solve the $Q$ integral of 3N-folds, the sense of integral is explained as followings. Usually, one-dimensional integral $I_{1D} = \int_a^b f(x)dx$ is interpreted as the sum of infinite number of rectangles with area $A_i = f(x_i)\Delta x$, i.e., $I_{1D} = \lim_{\Delta x \to 0} \sum_i A_i$. While from another angle, the length of the 1D element $\Delta x$ at $x_i$ is modulated by $f(x_i)$ to be a new length element $\Delta x'_i = f(x_i)\Delta x$ and $I_{1D} = \sum_i \Delta x'_i$. In other words, 1D integral is a summation of length elements instead of area elements and equals to an effective length of |b-a|. Similarly, a two-dimensional integral $I_{2D} = \int_0^a \int_0^b dx\, dy\, f(x,y)$ equals to an effective area of $a \cdot b$ because the area element $ds = dxdy$ is enlarged (or shrunk) by *f(x,y)* giving rise to an effective area element $ds' = f(x,y)dxdy$. Followed by this notion, a N-fold integral $I_{ND} = \int_0^{a_1} \int_0^{a_2} \ldots \int_0^{a_N} dx_1 dx_2 \ldots dr_N f(x_1, x_2 \ldots x_N)$ equals to an effective volume of $a_1 \cdot a_2 \ldots a_N$.

When the integrand $f(x_1, x_2 \ldots x_N)$ is in a form of exp[-$U(x_1, x_2 \ldots x_N)$] with $U(x_1, x_2 \ldots x_N)$ being positive definite within the entire integral domain and having minimum at the origin ($U(0) = 0$), the effective length of $a_i$ is defined as[25]

$$a'_i = \int_0^{a_i} \exp[-U(0 \ldots x_i \ldots 0)]dx_i, (i = 1,2 \ldots N) \quad (4)$$

and the effective volume approximates to a product $\prod_{i=1}^{N} a'_i$, i.e.,

$$I_{ND} \cong \prod_{i=1}^{N} a'_i. \quad (5)$$

For the 3*N*-folds integral of Eq. (3), although the integrand is of the same form as required by Eq. (5), it may not be positive definite or have no minimum at the origin. Letting the set $q^{3N} = \{q_1, q_2 \ldots q_{3N}\}$ be the coordinates of particles in state of the lowest potential energy $U_0$, we may introduce a function

$$U'(x'^{3N}, X^{3M}) = U(x^{3N}, X^{3M}) - U_0 \quad (6)$$

with $x_i' = x_i - q_i$. By inserting Eq. (6) into Eq. (3), we obtain

$$Q = e^{-\beta U_0} \int \exp[-\beta U'(x'^{3N}, X^{3M})] \qquad (7)$$

Clearly, $U'(x'^{3N}, X^{3M})$ is positive definite within all the integral domain and has minimum at to the origin ($U'(0, X^{3M}) = 0$). According to Eq. (5), the integral in Eq. (7) equals to an effective $3N$-dimensional volume,

$$Q = e^{-\beta U_0} \prod_{i=1}^{3N} L_i, \qquad (8)$$

where the effective length $L_i$ on the $i$th degree of freedom is defined as

$$L_i = \int e^{-\beta U'(0...x_i'...0, X^{3M})} dx_i' \qquad (9)$$

In this way, the $3N$-fold integral turns into one-fold integral.

For homogeneous 3D materials with one component, such as one component crystals, the effective length of an arbitrary atom in one degree of freedom (such as $L_x$) may be the same as the other two ($L_y$ and $L_z$) and equates to the ones of other atoms. For a 2D material sheet on substrate (Fig.1), however, the effective length $L_z$ might be different from $L_y$ or $L_x$, and the edged atoms ($N_1$) should have different effective lengths from the ones of the atoms ($N_2$) in the center region. In such case, Eq. (8) turns into,

$$Q = e^{-\beta U_0} [L_x^1 L_y^1 L_z^1]^{N1} \cdot [L_x^2 L_y^2 L_z^2]^{N2}. \qquad (10)$$

To obtain the effective lengths, the first step is to find the most stable structure of the 2D materials with the lowest potential $U_0$, which can be accomplished by a dynamic damping method[27, 28]. Starting from the most stable structure, one atom in the center region (or in the edged region) is moved step by step in one of the degree of freedom, such as X- axis, while the Y- and Z- coordinates and all other atoms keep fixed to determine $U'(0 \dots x'_i \dots 0, X^{3M})$ for calculating the $L_x^1, L_y^1, L_z^1$ (or $L_x^2, L_y^2, L_z^2$).

## 3. Applying to carbon 2D materials

The approach developed in the last section was demonstrated by calculating the PF for a piece of graphene (Fig. 2(a)) or γ-graphyne (Fig. 2(c)) sheet of 510 C atoms on the (111) surface of Cu substrate of 2640 atoms arranged in perfect fcc lattices. Brenner function[26] was employed for the C atoms interaction, and $V(x^{3N}, X^{3M})$ was taken as the summation of Lennard-Jones(L-J) function $f(r) = 4\varepsilon(\frac{\sigma^{12}}{r^{12}} - \frac{\sigma^6}{r^6})$ for pairwise interaction between a C and a Cu atom with ε=0.0168eV, $\sigma = 2.2$ Å[29].

The system was cooled below 0.01K by a damping method[30] to determine the lowest energy $U_0$ and the most stable structure. According to the configuration, the C atoms can be grouped as center or edged atoms, and for each of the atom with different surroundings one of its coordinate (such as $x'$) was changed step by step with an interval of 0.001 Å while its $y'$ and $z'$ coordinates and all other atoms were kept fixed to record the $U'(0 \ldots x'_i \ldots 0, X^{3M})$, as shown in Fig. 2 (b) and (d) for graphene and γ-graphyne, respectively. The $U'$ for the center atom is indeed different from that for the edged atoms, and for the same atom, the $U'$ along one coordinate axis is also different from the other two, so the configuration integral was conducted by

$$Q = e^{-\beta U_0}[L_X^1 L_Y^1 L_Z^1]^{N1} \cdot [L_X^2 L_Y^2 L_Z^2]^{N2} \cdot [L_X^3 L_Y^3 L_Z^3]^{N3} \qquad (11)$$

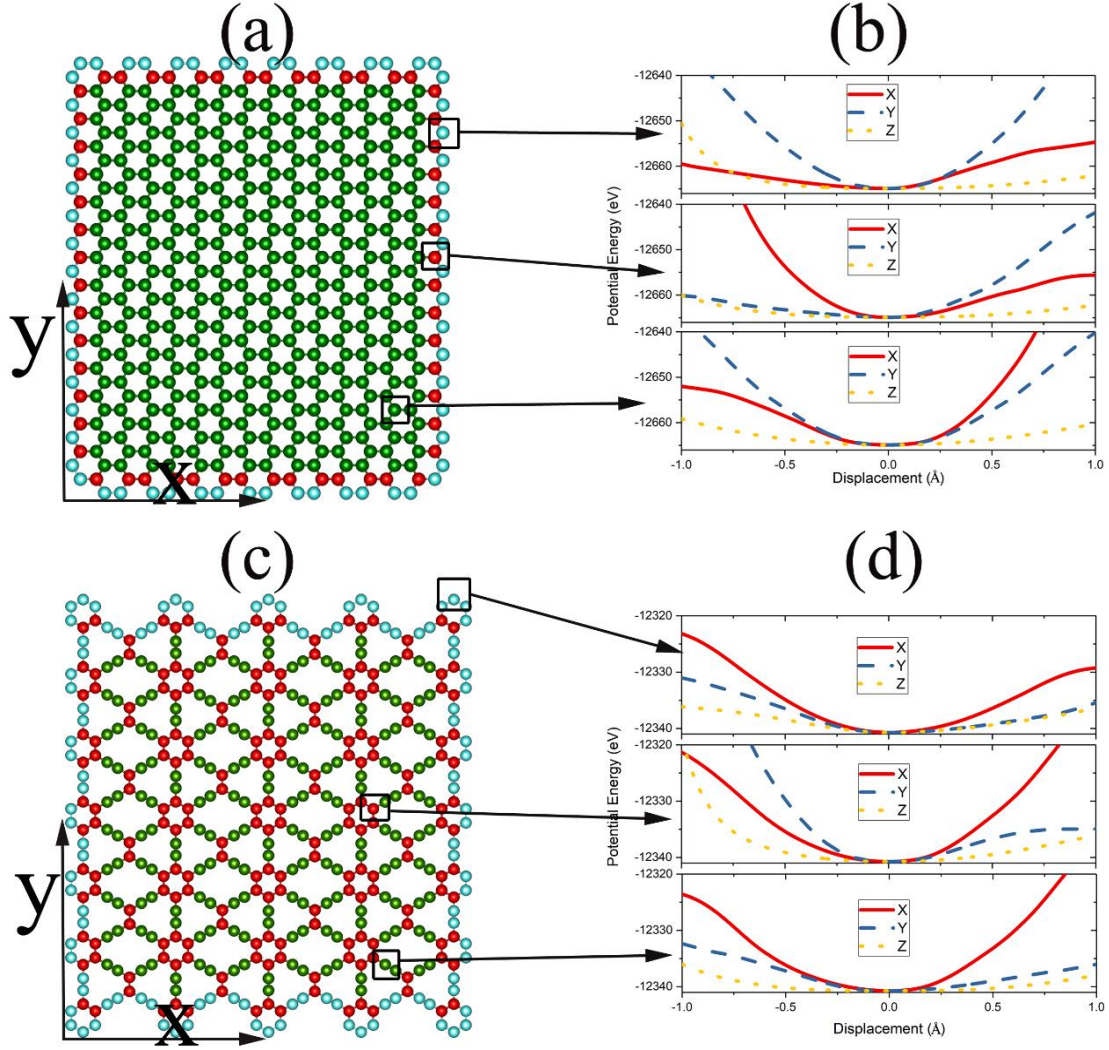

**Figure 2.** Top view of a graphene sheet (a) and a γ-graphyne sheet (c) on a Cu (111) substrate. The potential energy (b; d) felt by a C atom moving along the X-, Y- or Z-axis depends on the specific surrounding of the C atom.

Applying Eq. (2) and $E = -\frac{\partial}{\partial \beta}\ln Z$, the internal energy ($E_{PF}$) was obtained through

$$E_{PF} = \frac{3}{2}Nk_B T + \frac{k_B T^2}{Q}\frac{\Delta Q}{\Delta T} \qquad (12)$$

with $\Delta T = 0.1\text{K}$.

In order to test accuracy of the above algorithm, a common procedures for MD simulations of a canonical ensemble [28] was employed to produce the internal energy of the 2D materials contacted with a thermal bath at given temperature *T*, and the Verlet Algorithm was employed for integrating the equations of motions with time step 0.2 *fs*. Within the first 400 *fs*, all the carbon atoms are assigned velocities every 40 *fs* according to the Maxwell velocity distribution at temperature *T*, and then the internal energy ($E_{MD}$) and the temperature were recorded every 30 *fs* to do the average over 100 records.

As shown in Fig. 3, the internal energy ($E_{PF}$) derived from the PF is in excellent agreement with that ($E_{MD}$) obtained from the MD simulations. In the temperature range from 100K to 1300K, the relative error ($\frac{|E_{PF}-E_{MD}|}{|E_{MD}|} \times 100\%$) is below 0.03%, and only 0.0005% and 0.002% for the graphene at 500K and γ-graphyne at 1100K, respectively. It is notable that the dependence of internal energy (*E*) on temperature is nearly linear, indicating that $E = U_0 + BNk_B T$, where *B* is a constant. According to statistical physics, the constant $B$ equals to 3 for 3D crystal atoms with harmonic coordinates. However, for the graphene and γ-graphyne sheet the constant $B$ equals to 2.97 and 2.90, respectively, implying that the C atoms are not in the motion of harmonics.

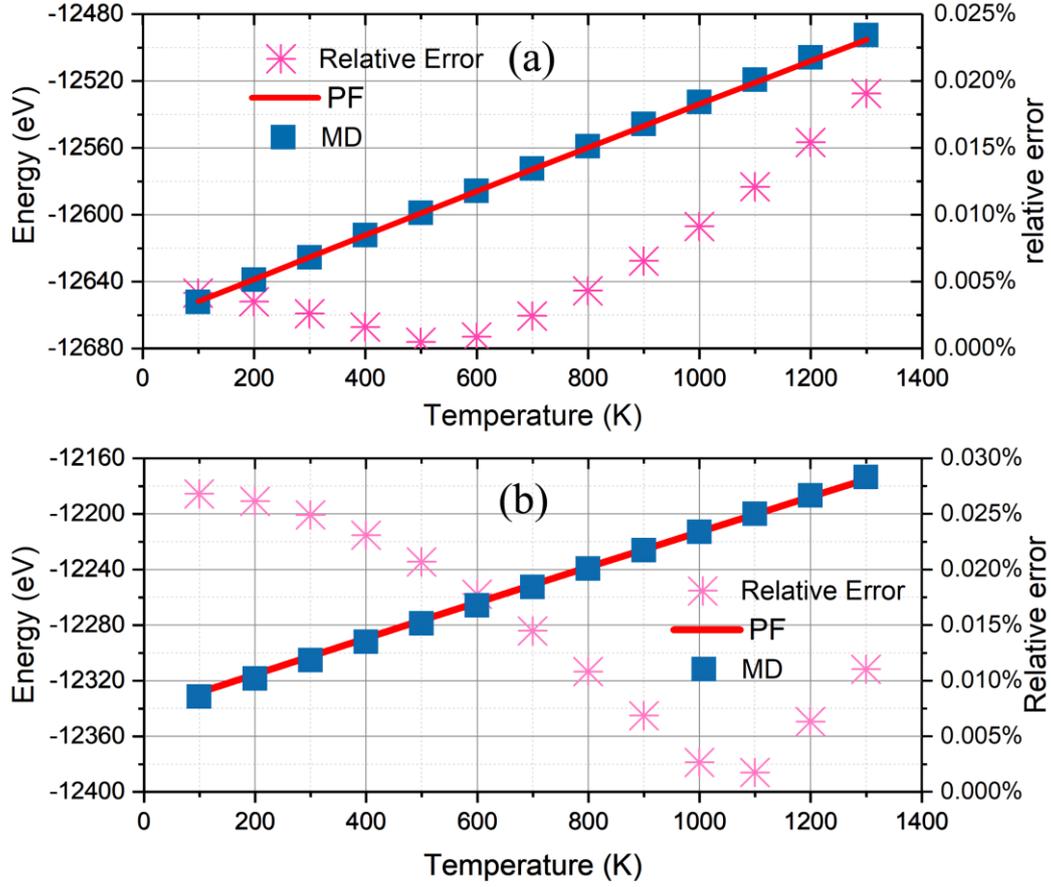

**Figure 3.** The internal energy as the function of temperature derived from the PF (red line) or MD simulation (blue square) for the graphene sheet (a) and the γ-graphyne sheet (b), where the relative errors $\frac{|E_{PF}-E_{MD}|}{|E_{MD}|} \times 100\%$ (pink stars) are characterized at the right vertical axis.

As for the calculation efficiency of our method, we would like to compare it with the ANS [24] assumed to be applicable to 2D materials. Usually, ANS must run at least one thousand steps to produce the partitions of the total potential energy, and in each step more than $3 \times 10^4$ configurations should be randomly produced for calculating the total potential energy, i.e., the times of the energy calculation is more than $3 \times 10^7$. Using our method, the times of the energy calculations is only $1.8 \times 10^4$, implying that it works at least 1000 times faster than the ANS.

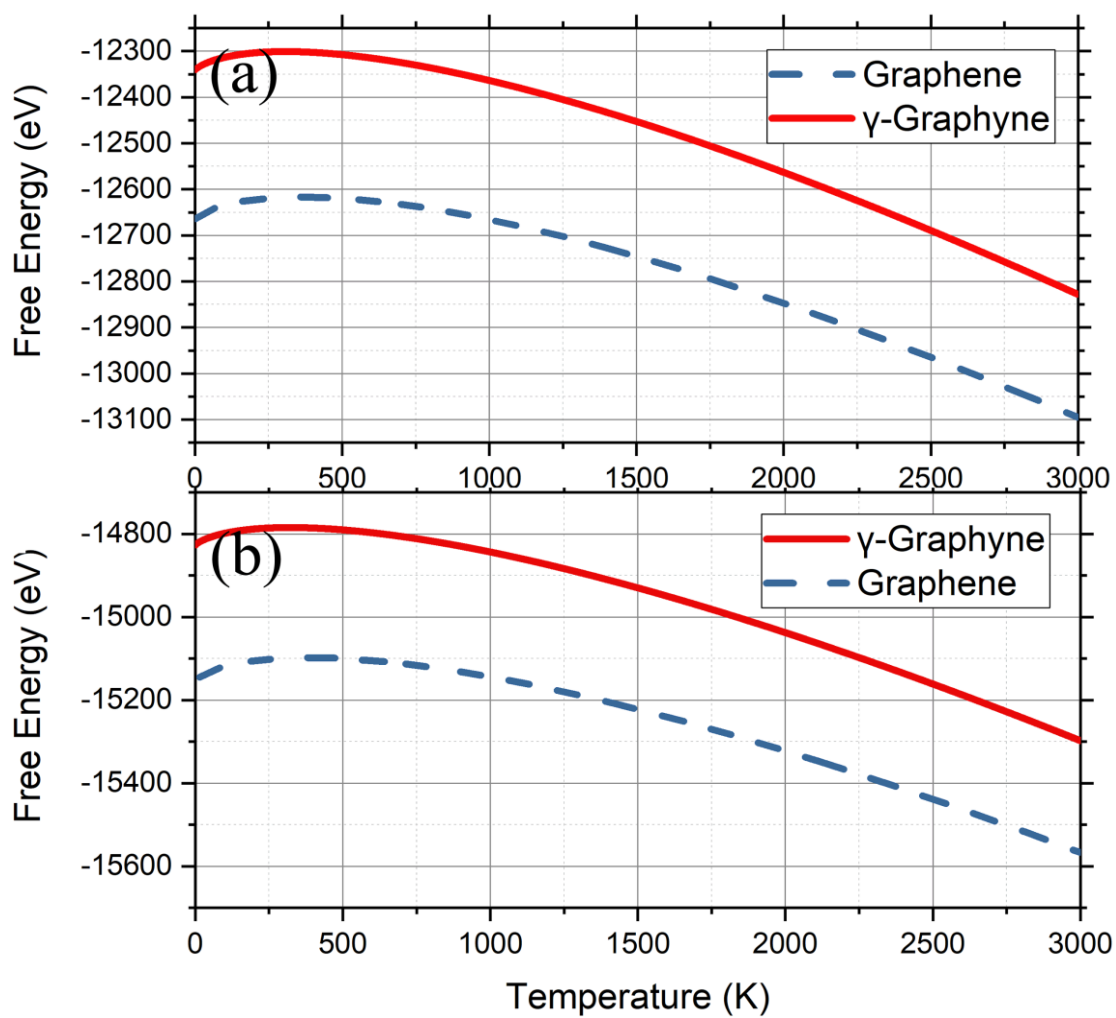

**Figure 4**. Free energy derived from the PF of graphene (blue dashed lines) and γ-graphyne (red solid lines) on the substrates of Cu [111] (a) and Ni [111] (b).

Applying $F = -k_B T \ln Z$, the free energy of graphene and γ-graphyne on Cu (111) was calculated (Fig. 4(a)), showing that the free energy of graphene is always smaller than the one of γ-graphyne in the temperature range from 0 to 3000K. The difference at 0K, 324 eV, decreases gradually down to 267 eV with the temperature up to 3000K, indicating that graphene should be more easily grown than γ-graphyne on Cu (111) surface via deposition of C atoms. We also calculated the free energy on Ni (111) surface (Fig. 4(b)) using the same method and found that graphene still owns free energy smaller than γ-graphyne although the difference at 0K, 319 eV gets gradually small with the temperatures. These results are consistent with previous experimental observations [31, 32].

## 4. Summary

In summary, an approach was developed to calculate the free energy (or partition function) of 2D material on substrate and was validated by MD simulations. The approach works at least 3 orders faster than ANS method and lays a solid foundation to precisely predict the thermodynamic properties and optimal growth conditions for novel 2D materials.

## 5. Acknowledgement

The work is supported by National Natural Science Foundation of China under Grant No.21727801 and No.11274073.

# References

[1] Novoselov KS, Geim AK, Morozov SV, Jiang D, Zhang Y, Dubonos SV, et al. Electric field effect in atomically thin carbon films. Science. 2004;306(5696):666-9.

[2] Li GX, Li YL, Liu HB, Guo YB, Li YJ, Zhu DB. Architecture of graphdiyne nanoscale films. Chem Commun. 2010;46(19):3256-8.

[3] Vogt P, De Padova P, Quaresima C, Avila J, Frantzeskakis E, Asensio MC, et al. Silicene: Compelling Experimental Evidence for Graphenelike Two-Dimensional Silicon. Physical Review Letters. 2012;108(15):155501.

[4] Yuhara J, Shimazu H, Ito K, Ohta A, Araidai M, Kurosawa M, et al. Germanene Epitaxial Growth by Segregation through Ag(111) Thin Films on Ge(111). ACS Nano. 2018;12(11):11632-7.

[5] Liu Z, Song L, Zhao S, Huang J, Ma L, Zhang J, et al. Direct Growth of Graphene/Hexagonal Boron Nitride Stacked Layers. Nano Letters. 2011;11(5):2032-7.

[6] Pakdel A, Zhi C, Bando Y, Golberg D. Low-dimensional boron nitride nanomaterials. Materials Today. 2012;15(6):256-65.

[7] Kaur H, Yadav S, Srivastava AK, Singh N, Schneider JJ, Sinha OP, et al. Large Area Fabrication of Semiconducting Phosphorene by Langmuir-Blodgett Assembly. Scientific reports. 2016;6:34095-.

[8] Reis F, Li G, Dudy L, Bauernfeind M, Glass S, Hanke W, et al. Bismuthene on a SiC substrate: A candidate for a high-temperature quantum spin Hall material. Science. 2017;357(6348):287-90.

[9] Bosi M. Growth and synthesis of mono and few-layers transition metal dichalcogenides by vapour techniques: a review. RSC Adv. 2015;5(92):75500-18.

[10] Ghidiu M, Lukatskaya MR, Zhao M-Q, Gogotsi Y, Barsoum MW. Conductive two-dimensional titanium carbide 'clay' with high volumetric capacitance. Nature. 2014;516:78.

[11] Benameur MM, Radisavljevic B, Héron JS, Sahoo S, Berger H, Kis A. Visibility of dichalcogenide nanolayers. Nanotechnology. 2011;22(12):125706.

[12] Bointon T, Barnes M, Russo S, Craciun M. High Quality Monolayer Graphene Synthesized by Resistive Heating Cold Wall Chemical Vapor Deposition; 2015.

[13] Leandri C, Oughaddou H, Aufray B, Gay JM, Le Lay G, Ranguis A, et al. Growth of Si nanostructures on Ag(001). Surface Science. 2007;601(1):262-7.

[14] Aufray B, Kara A, Vizzini S, Oughaddou H, Leandri C, Ealet B, et al. Graphene-like silicon nanoribbons on Ag(110): A possible formation of silicene. Appl Phys Lett. 2010;96(18):3.

[15] Lalmi B, Oughaddou H, Enriquez H, Kara A, Vizzini S, Ealet B, et al. Epitaxial growth of a silicene sheet. Appl Phys Lett. 2010;97(22):2.

[16] Lazic P, Crljen Z. Graphyne on metallic surfaces: A density functional theory study. Physical Review B. 2015;91(12):5.

[17] Yuan QH, Ding F. Formation of carbyne and graphyne on transition metal surfaces. Nanoscale. 2014;6(21):12727-31.

[18] Tang Y, Yang H, Yang P. Investigation on the contact between graphdiyne and Cu (111) surface. Carbon. 2017;117:246-51.

[19] Swendsen RH, Wang J-S. Replica Monte Carlo Simulation of Spin-Glasses. Physical Review Letters. 1986;57(21):2607-9.

[20] Bartels C. Analyzing biased Monte Carlo and molecular dynamics simulations. Chemical Physics Letters. 2000;331(5):446-54.

[21] Laio A, Parrinello M. Escaping free-energy minima. Proceedings of the National Academy of

Sciences. 2002;99(20):12562.

[22] Wang F, Landau DP. Efficient, Multiple-Range Random Walk Algorithm to Calculate the Density of States. Physical Review Letters. 2001;86(10):2050-3.

[23] Pártay LB, Bartók AP, Csányi G. Efficient Sampling of Atomic Configurational Spaces. The Journal of Physical Chemistry B. 2010;114(32):10502-12.

[24] Do H, Wheatley RJ. Density of States Partitioning Method for Calculating the Free Energy of Solids. Journal of Chemical Theory and Computation. 2013;9(1):165-71.

[25] Ning B-Y, Gong L-C, Weng T-C, Ning X-J. Solution of partition function for macroscopic condensed matters-a long standing key problem in statistical physics. arXiv:190108233.

[26] Brenner DW. EMPIRICAL POTENTIAL FOR HYDROCARBONS FOR USE IN SIMULATING THE CHEMICAL VAPOR-DEPOSITION OF DIAMOND FILMS. Physical Review B. 1990;42(15):9458-71.

[27] Zhang Q, Buch V. COMPUTATIONAL STUDY OF FORMATION DYNAMICS AND STRUCTURE OF AMORPHOUS ICE CONDENSATES. J Chem Phys. 1990;92(8):5004-16.

[28] Ye XX, Ming C, Hu YC, Ning XJ. Evaluating the ability to form single crystal. J Chem Phys. 2009;130(16):6.

[29] Shi XH, Yin QF, Wei YJ. A theoretical analysis of the surface dependent binding, peeling and folding of graphene on single crystal copper. Carbon. 2012;50(8):3055-63.

[30] Ning XJ, Qin QZ. A new molecular dynamics method for simulating trapping site structures in cryogenic matrices. J Chem Phys. 1999;110(10):4920-8.

[31] Yu Q, Jauregui LA, Wu W, Colby R, Tian J, Su Z, et al. Control and characterization of individual grains and grain boundaries in graphene grown by chemical vapour deposition. Nat Mater. 2011;10(6):443-9.

[32] Li X, Cai W, Colombo L, Ruoff RS. Evolution of Graphene Growth on Ni and Cu by Carbon Isotope Labeling. Nano Letters. 2009;9(12):4268-72.